\documentclass{mn2e}
\usepackage{psfig}

\newcommand\etal{et al. }

\newcommand\xmm{{\it XMM-Newton}}

\newcommand\asca{{\it ASCA}}

\newcommand\nh{N_{H}}

\begin{document}

   \title{The origin of the Fe K features in Markarian~205 and 
Markarian~509}
   \author[Page et al]{M.J. Page$^{1}$, S.W. Davis$^{2}$, N.J. Salvi$^{1}$
\\
$^{1}$Mullard Space Science Laboratory, University College London,
Holmbury St Mary, Dorking, Surrey, RH5 6NT, UK\\
$^{2}$Department of Physics, University of California, Santa Barbara, CA 93106,
USA\\
}

\maketitle
\begin{abstract}
We examine the 3-10 keV EPIC spectra of Mrk~205 and Mrk~509 to investigate their
Fe K features. The most significant feature in the
spectra of both objects is an emission line at 6.4 keV.  
The spectra can be adequately modelled
with a power law and a relatively narrow ($\sigma < 0.2$ keV) Fe~K${\alpha}$
emission line. Better fits are obtained when an additional Gaussian emission 
line,
relativistic accretion-disk line, or Compton reflection from cold material, is
added to the spectral model. We obtain similar goodness of fit for any of these
three models, but the model including Compton reflection from cold
material offers the simplest, physically self-consistent solution, because it
only requires one reprocessing region.
Thus the Fe K spectral features in Mrk~205 and Mrk~509 do not present
strong evidence for reprocessing in the inner, relativistic parts of
accretion disks.

\end{abstract}

%

\maketitle

\section{Introduction}
\label{sec:introduction}

X-ray observations probe the central regions of AGN. In the standard
paradigm, this corresponds to the inner parts of 
an accretion disk around a supermassive
black hole. Above the disk, a hot corona Compton upscatters
optical-EUV photons to X-ray energies; some of this X-ray radiation is
reprocessed in the surrounding material including the disk, giving
rise to prominent Fe K$\alpha$ lines. The broad, distorted velocity 
profile of Fe
K$\alpha$ emission, suggesting an accretion disk around a supermasive black 
hole, was 
 first observed with \asca\ 
in MCG --6-30-16
(Tanaka \etal \nocite{tanaka95} 1995). This profile, with a sharp blue wing 
and 
a broad 
red tail, is a remarkable 
probe of the strong gravity regime.

Studies of other AGN with \asca\ suggested that broad, low-ionisation
Fe K$\alpha$ emission is common in AGN (Nandra \etal \nocite{nandra97}
1997, Reynolds \etal \nocite{reynolds97} 1997), but it was not until
the launch of \xmm\ that the diversity of Fe line profiles could
really be investigated. With the large increase in collecting area
afforded by \xmm, it was soon noticed that some luminous AGN showed
different Fe line profiles to that of MCG --6-30-15 (e.g. Reeves \etal
\nocite{reeves01a} 2001a, Blustin \etal \nocite{blustin02} 2002). One
particularly interesting example, the
luminous Seyfert 1 galaxy Mrk~205, appeared to have a narrow, neutral
Fe K$\alpha$ line at 6.4 keV, accompanied by a broad line from He-like
Fe (Reeves \etal \nocite{reeves01b} 2001b). Reeves \etal argued that the
ionised Fe line originates in the inner parts of an accretion disk, while the
neutral Fe K$\alpha$ line originates in a molecular torus,
hypothesised to lie outside the broad line regions in AGN unification
schemes (Antonucci \nocite{antonucci93} 1993). 
A later observation of another Seyfert galaxy, Mrk~509, showed
a very similar pair of narrow-neutral and broad-ionised Fe K$\alpha$
emission lines (Pounds \etal \nocite{pounds01} 2001), demonstrating that this  
configuration of Fe K$\alpha$ profiles is not an isolated phenomenon in
Mrk~205.

In this paper we revisit the Fe K features in Mrk~205 and Mrk~509 seen by
\xmm. By coadding the spectra from the three EPIC instruments we are able to
maximise the signal to noise per bin while properly sampling the EPIC spectral
resolution around Fe K. The paper is laid out as follows.  In Section
\ref{sec:observation} we describe the observations and data reduction, and we
describe the spectral fitting in Section \ref{sec:results}. The results are
discussed in Section \ref{sec:discussion} and we present our conclusions in
Section \ref{sec:conclusions}. The Appendix contains a description of the
method employed to coadd spectra from the different EPIC instruments.

\section{Observations and data reduction}
\label{sec:observation}

Mrk~205 was observed with \xmm\ on the 7th May 2000, and these data were
presented by Reeves \etal \nocite{reeves01b} (2001b). Several exposures were
taken in both MOS and PN cameras, in full frame and large-window modes.
Spectra of the source were extracted from circular regions of radius $\sim
50''$ and background spectra were obtained from nearby source-free regions.
All valid event patterns (singles, doubles, and triples) were selected in MOS,
and only single events in PN. The spectra were combined using the procedure
outlined in Appendix A.

Mrk~509 has been observed twice with \xmm. The first observation took place on
25th November 2000 and the data were presented by Pounds 
\etal \nocite{pounds01}
(2001). The second observation was performed on the 20th April 2001. In both
observations, MOS and PN cameras were operated in small window mode.
Source spectra were taken from a circular region of $40'' - 50''$ radius, and
background spectra were obtained from nearby regions free from bright sources.
Single events were selected in MOS,
and single and double events were selected in PN. The spectra were 
combined using the procedure
outlined in the Appendix to produce one spectrum for each
observation and one spectrum which is a combination of the two. 

\begin{table}
\caption{\xmm\ observations of Mrk~205 and Mrk~509}
\label{tab:observations}
\begin{tabular}{lccc}
Object&Date&Exposure &Count rate \\
&&(ks)&(count s$^{-1}$)\\
Mrk~205 & 7 May 2000 & 49.0 & 4.9\\
Mrk~509 & 25 November 2000 & 23.4 & 26.7\\
Mrk~509 & 20 April 2001 & 30.6 & 38.3\\
\hline
&&&\\
\end{tabular}
\end{table}

\section{Results}
\label{sec:results}

\begin{table*}
\caption{Spectral fits. Errors are quoted at 95\% confidence for 1 interesting
parameter. In cases for which a parameters has reached a limit on the allowed
range of values while $\Delta \chi^{2} < 4$ the limit has been marked with a
`*'. Parameters with values that are fixed in the fit are marked '$F$'.}
\label{tab:results}
\begin{center}
\leavevmode
\vspace{-20mm}
\psfig{figure=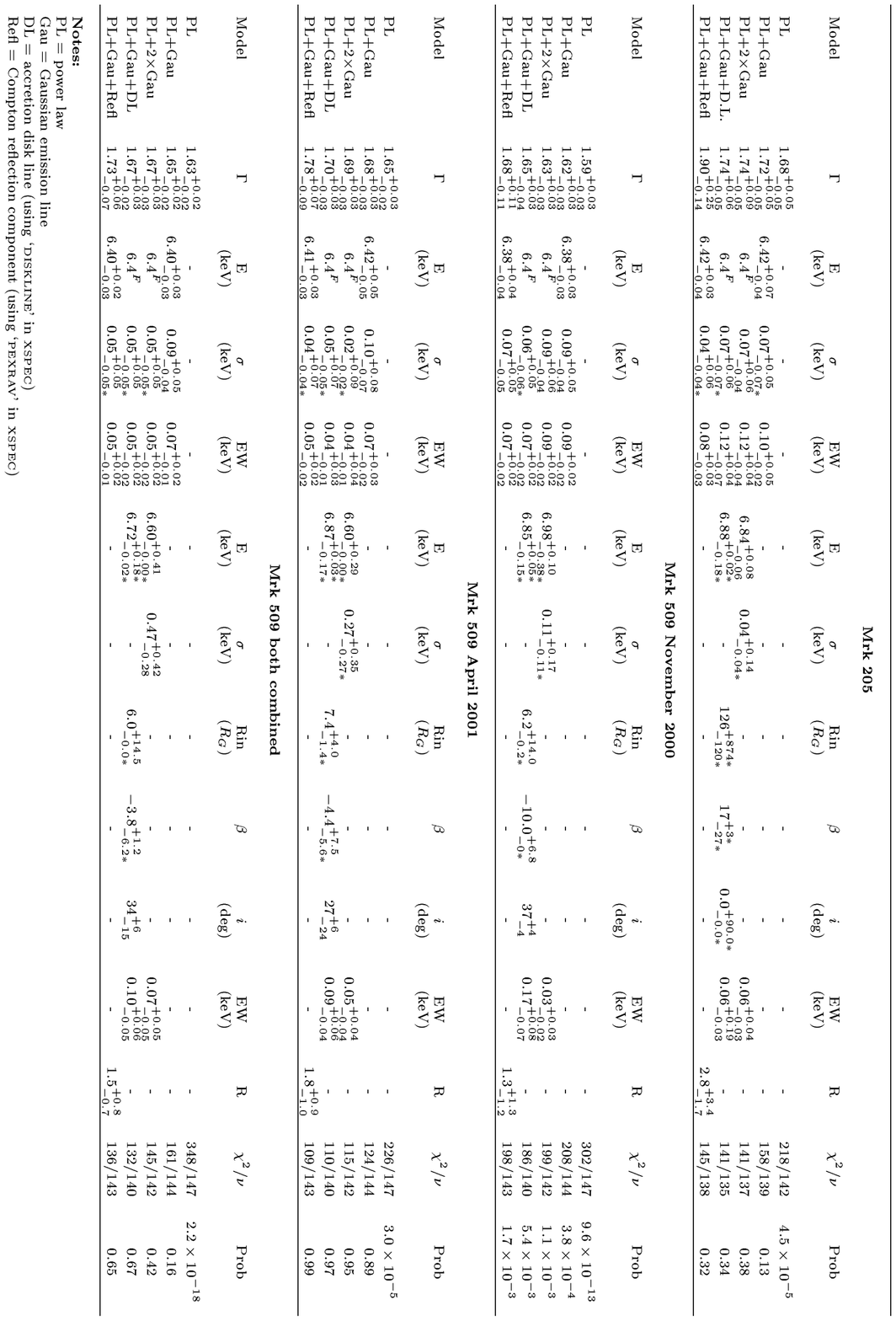,width=150truemm}
\vspace{20mm}
\end{center}
\end{table*}

The spectral fitting was performed with {\small XSPEC}. Only the rest-frame
3-10 keV energy range was used in the spectral fitting because we are
primarily interested in the Fe K features. 
The broad emission lines reported
by Reeves \etal \nocite{reeves01b} (2001b) and
Pounds \etal \nocite{pounds01} (2001) are only significant between 5 and 8 keV
(see Fig. 4 of Pounds \etal \nocite{pounds01} 2001), so the 3-10 keV 
energy range allows a good
measurement of the continuum on either side of Fe K, while excluding the
noisier and less-well calibrated data at higher energies and the complex 
spectrum found at lower
energies.
We included the small effect of absorption from the Galaxy as a component in 
all our spectral modelling
($\nh = 2.9 \times 10^{20}$ cm$^{-2}$ towards Mrk~205 and 
$\nh = 4.1 \times 10^{20}$ cm$^{-2}$ towards Mrk~509, Dickey \& Lockman
\nocite{dickey90} 1990).
The results of the spectral fits are
given in Table \ref{tab:results}.

\subsection{Mrk~205}

\begin{figure}
\begin{center}
\leavevmode
\psfig{figure=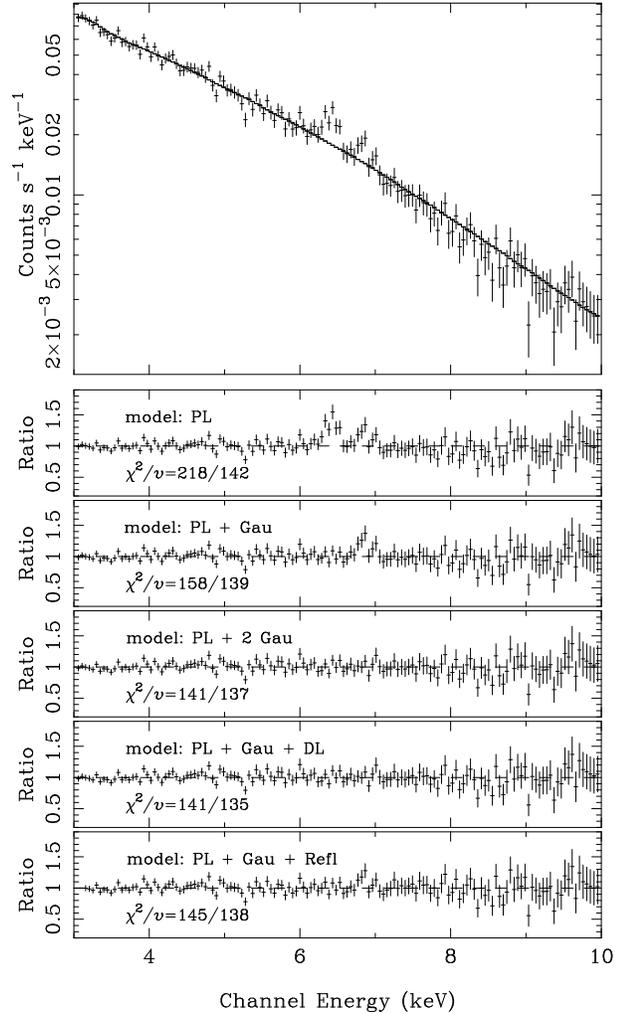,width=80truemm}
\caption{3-10 keV EPIC spectrum of Mrk~205 and power law model (top panel) and
data/model ratio for the five different model fits (lower panels).}
\label{fig:mrk205_allspec}
\end{center}
\end{figure}

We began by fitting a power law model. The counts spectrum is shown in
Fig. \ref{fig:mrk205_allspec}, along with the power law model convolved with
the instrument response.  Like Reeves \etal \nocite{reeves01b} (2001b)
we find this model is a poor fit, and there are significant residuals
around 6.4 keV.  We therefore added a gaussian emission line, and
obtained an acceptable fit, with a resolved line consistent with 6.4
keV Neutral Fe K$\alpha$. Although the fit was acceptable, residuals
remained at $\sim 7$ keV, and so we tried further fits including one of 
three additional model
components that might plausibly account for these residuals and
improve the fit: a gaussian emission line at $E>6.5$ keV, a
relativistically-broadened accretion disk line (Fabian \etal
\nocite{fabian89} 1989; the ``diskline'' model in {\small XSPEC}), 
and Compton reflection from cold material
(``pexrav'' in {\small XSPEC}; Magdziarz and Zdziarski \nocite{magdziarz95}
1995). In the diskline model the energy of the line was constrained to
lie between 6.7 and 6.9 keV, corresponding to He-like or H-like Fe (as
proposed by Reeves \etal \nocite{reeves01b} 2001b). In the reflection
model the inclination was fixed at 45$^{o}$, and we assumed Solar
elemental abundances (Anders and Grevesse \nocite{anders89} 1989). The
addition of any of these three components resulted in a similar
goodness of fit. The energies of the second Gaussian line and diskline
components are found to be consistent with Fe~XXVI, and in the case of
the diskline the best fit was found for a line produced many $R_{G}$
from the black hole giving it a narrow profile, similar to that of the
Gaussian line. We have also tried more complex models, including two emission 
lines
as well as reflection: the lowest value of $\chi^{2}/\nu=138/135$ was obtained
when the second emission line is a Gaussian. However, according to the F-test,
this is only a 1$\sigma$ improvement over the model including a single emission
line and Compton reflection. 

\subsection{Mrk~509 first observation}

\begin{figure}
\begin{center}
\leavevmode
\psfig{figure=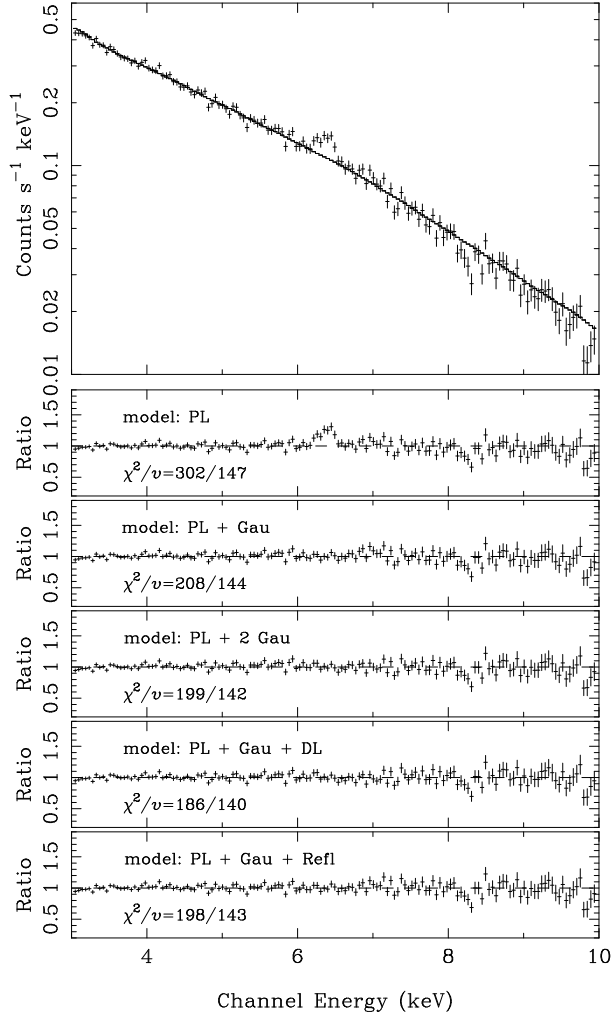,width=80truemm}
\caption{3-10 keV EPIC spectrum of Mrk~509 from the first observation
and power law model (top panel) and
data/model ratio for the five different model fits (lower panels).}
\label{fig:mrk509_1_allspec}
\end{center}
\end{figure}

As for Mrk~205, we began with a power law model and obtained a
completely unacceptable fit with large residuals around 6.4
keV (see Fig. \ref{fig:mrk509_1_allspec}). 
Addition of a Gaussian line at $\sim 6.4$ keV resulted in a much better, 
but still poor fit (rejected with $> 99.9\%$ confidence). 
Adding a further Gaussian, a diskline or a reflection
component improved the goodness of fit, which is marginally better when 
the third model component is a diskline rather than a reflection
component or a second Gaussian line. However, whichever additional model 
component
is included the model is
still unacceptable at $>99\%$ confidence. 
We have also tested a more complex model, 
 including a second emission line as well as
 reflection, and obtain a best 
$\chi^{2}/\nu=182/138$ for a diskline; this is only marginally
better (2$\sigma$ according to the F-test) than the fit with reflection but
without the second emission line.
Strong residuals at 8.3 and 9.8 keV
together contribute $\sim 30$ to the $\chi^{2}$, and are therefore largely 
responsible for
the poor model fit. We have no credible explanation for these residuals other
than as statistical fluctuations.

\subsection{Mrk~509 second observation}

\begin{figure}
\begin{center}
\leavevmode
\psfig{figure=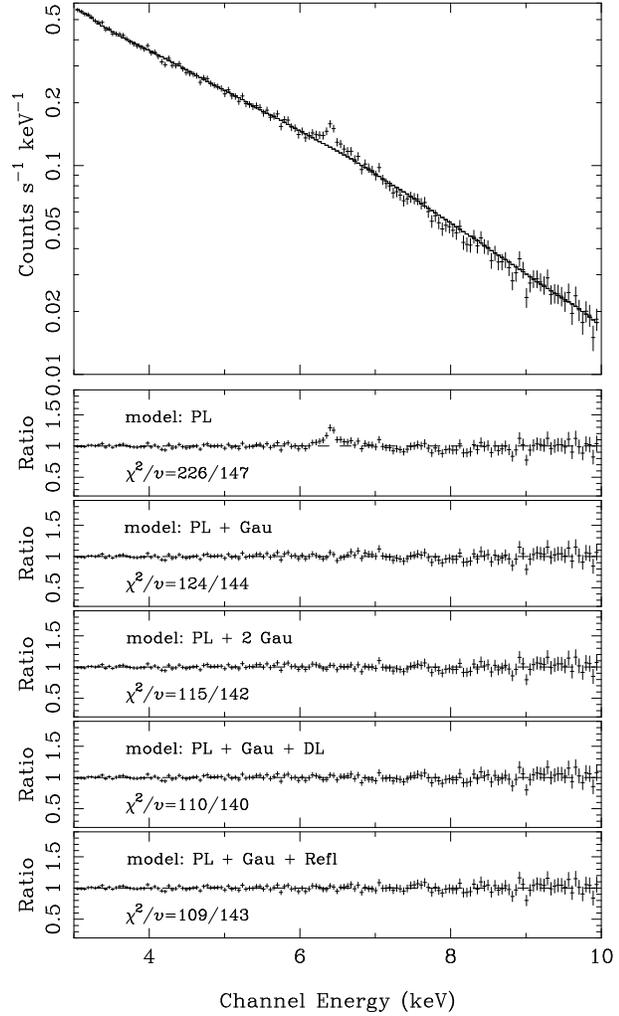,width=80truemm}
\caption{3-10 keV EPIC spectrum of Mrk~509 from the second observation 
and power law model (top panel) and
data/model ratio for the five different model fits (lower panels).}
\label{fig:mrk509_2_allspec}
\end{center}
\end{figure}

A power law model provided a poor fit to the spectrum 
(see Fig. \ref{fig:mrk509_2_allspec}), and once again we
found that the addition of a $\sim 6.4$ keV emission line resulted in
a significantly improved (and quite acceptable) $\chi^{2}/\nu$=124/144. 
The addition of an extra Gaussian, diskline or reflection component each 
improved the 
fit further, resulting in very a good fit. However, the addition of a second
emission line to the reflection model results in no further improvement in
$\chi^{2}/\nu$.

\subsection{Mrk~509 both observations combined}

\begin{figure}
\begin{center}
\leavevmode
\psfig{figure=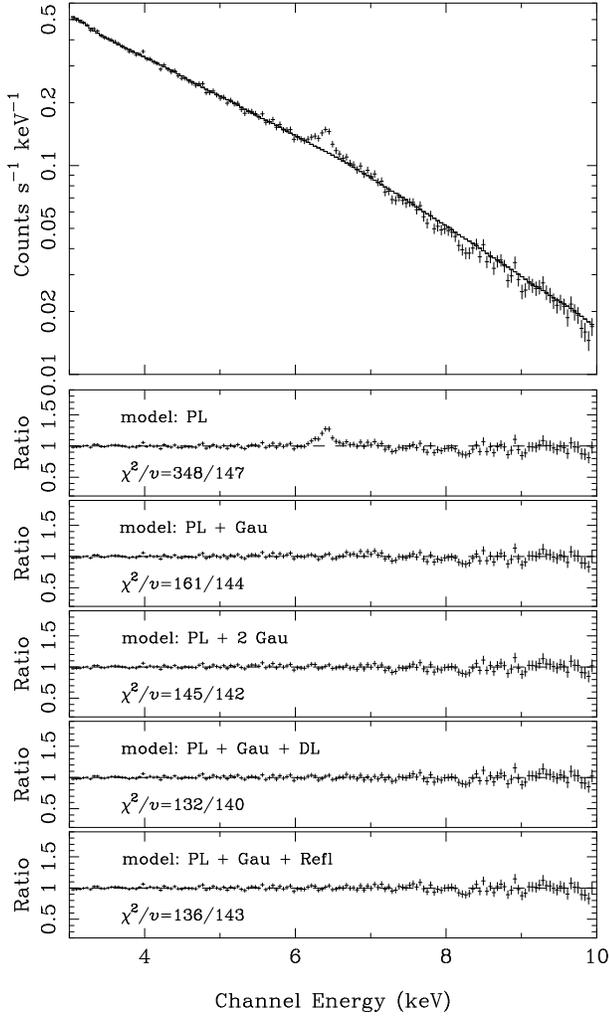,width=80truemm}
\caption{3-10 keV EPIC spectrum of Mrk~509 (both observations coadded) 
and power law model (top panel), and
data/model ratio for the five different model fits (lower panels).}
\label{fig:mrk509_allspec}
\end{center}
\end{figure}

The overall spectral shape of Mrk~509 is extremely similar in the two
observations, and so to improve signal to noise we have coadded the data from
both observations. A model containing only a power law is convincingly
rejected, but a power law and a single Gaussian
provide an acceptable fit to the data.  Adding an additional Gaussian, diskline
or reflection component makes a significant improvement to the $\chi^{2}/\nu$ 
(at $>99.9\%$ significance according to the F-test), resulting in a good
fit to the data. Finally, we have tried more complex models combining
reflection with a 
second Gaussian or accretion-disc emission line. In this case the 
$\chi^{2}/\nu$ is poorer than for the model including reflection without a
second emission line.

\section{Discussion}
\label{sec:discussion}

The most significant spectral feature in all the 3-10 keV EPIC spectra of
Mrk~205 and Mrk~509 is low-ionization Fe~K$\alpha$ emission at 6.4 keV.
In both sources, improved fits are obtained when an additional emission
component is included, peaking at slightly higher energy; Gaussian or
relativistic emission lines or cold reflection are all plausible forms for this
additional spectral component.
For Mrk~509, 
the spectral
modelling requires that the 6.4 keV line is broad, FWHM~$> 5000$~km~s$^{-1}$
unless this additional component is included. More complex models, combining a
second emission line with cold reflection, do not produce significantly better
fits for any of the spectra than models with cold reflection but without a 
second emission line.

The 6.4 keV emission is a
signature of reprocessing by cold material, but where does this
reprocessing take place?
Both galaxies show
negligible intrinsic absorption in soft X--rays, 
and hence the material
responsible for the 6.4 keV emission must lie outside the line of sight
to the continuum source.
Possible locations include the accretion disk, the molecular torus
favoured by AGN unification schemes (Antonucci \nocite{antonucci93} 1993), and
the (optical) broad line clouds. The molecular torus and the 
accretion disk represent Compton thick targets, and reprocessing at these
locations will result in a Compton reflection component with an edge at
7.1 keV as well as Fe K$\alpha$ line emission (Matt, Perola \& Piro
\nocite{matt91} 1991, George and Fabian \nocite{george91} 1991). However, the
broad line clouds are expected to be Compton thin (Shields, Ferland \&
Peterson \nocite{shields95} 1995) 
and therefore the Fe K$\alpha$ line emission will not be accompanied by
significant Compton scattered continuum. 

Simulations by Leahy and Creighton \nocite{leahy93} (1993, see also Yaqoob
\etal \nocite{yaqoob01} 2001) show that if the 
broad-line clouds have column densities of $10^{23}$~cm$^{-2}$, they would 
need to cover
50\% of the sky, as seen by the continuum source, to
produce the Fe K$\alpha$ line of 50 eV equivalent width in Mrk~509.
To produce the
100 eV equivalent width line seen in Mrk~205, the broad line clouds 
would need to surround $\sim 100\%$ of the central source. 
These covering fractions are much higher than the typical broad line region 
covering fractions deduced from the ultraviolet (10\% -- 25\%, Davidson \&
Netzer \nocite{davidson79} 1979, Goad \& Koratkar \nocite{goad98} 1998),
suggesting that the broad line regions are probably not responsible for the 
majority of
the Fe K$\alpha$ photons. Furthermore in Mrk~509, even if the broad line 
region does
produce the Fe K$\alpha$ line, something else must contribute an additional 
broad spectral
component at slightly higher energy, or else the velocity width of the Fe K$\alpha$ line is inconsistent with the
width of the
optical lines, which have FWHM of only 2270 km s$^{-1}$ (Wandel, Peterson \&
Malkan \nocite{wandel99} 1999).

On the other hand, if the Fe K$\alpha$ line originates in Compton thick
material, the equivalent width of the line suggests that this material 
intercepts $\sim$ 30 -- 60 per cent of the emitted radiation in Mrk~205,
and about half as much in Mrk~509 (George \& Fabian \nocite{george91}
1991). Therefore in both AGN the strength of the Fe K$\alpha$ line alone 
suggests 
that a
significant Compton reflection
component should be present.
When reflection is included in the fit we find
 that $ > 55\%$ of the radiation is intercepted by the reflector in
Mrk~205, and $ > 40\%$ in Mrk~509, slightly higher than inferred from the Fe
K$\alpha$ line but not greatly so. Thus a Compton thick reprocessor that
intercepts a significant fraction of the primary X-rays can account for all the
Fe K features in Mrk~205 or Mrk~509. This reprocessor could be a distant
molecular torus, or the accretion disk itself. However, the  
velocity width of the Fe K$\alpha$ line implies that even if the
reprocessor is the accretion disk, little of the reprocessed emission comes 
from the inner,
relativistic, parts of the disk.

So what evidence do we have for emission from highly ionized Fe, potentially in
an accretion disk? For Mrk~205, the fits with a second emission line are as
good as those including a reflection component. However the second emission
line is narrow when it is modeled as a Gaussian, and similarly the best fit
accretion disk line is one in which the emission is dominated from material in
the outer disk, resulting in a narrow line. Hence even if the spectrum of
Mrk~205 {\em does} include Fe~XXV or Fe~XXVI, there is no evidence for
relativistic broadening. 

For Mrk~509, the best-fit second emission line is
broad, whether it is modeled as a Gaussian or as a disk line, and the
steep emissivity index implies that most of the emission would come from the
inner part of the disk. The disk would have an inclination of between 20 and 40
degrees, in agreement with the findings of Pounds \etal \nocite{pounds01} 
(2001). However, the $\chi^{2}/\nu$ for the fits including an accretion disk
line are only 
slightly better than the fits including
Compton reflection for the first observation of Mrk~509, and are slightly
poorer for the second
observation;
the combined spectrum is as well fit by either model.
Hence although the features in the 
spectrum of
Mrk~509 can be fit with a model including both a distant, cold,
Compton-thin reprocessor and relativistically broadened
emission from highly ionised Fe in an accretion disk, they can be fit
equally well by reprocessing and Compton reflection 
from distant, cold 
material. 

Thus in both Mrk~205 and Mrk~509, we find that the Fe~K features can
be explained by a single phase Compton thick cold reflector. While the presence
of reflection from the highly ionized, high velocity, inner parts of an
accretion disk is not ruled out by these data, it is not unambiguously 
detected.

\section{Conclusions}
\label{sec:conclusions}

We have analysed the 3-10 keV \xmm\ EPIC spectra of Mrk~205 and Mrk~509 to
investigate the Fe K features in these objects. Acceptable fits can be obtained
for models containing nothing more than a power law and an emission line at 6.4
keV, consistent with cold Fe K$\alpha$. However, better fits are obtained when
an additional spectral component is included in the model, either Compton 
reflection from cold material or an emission line from ionised Fe; the goodness
of fit is similar whichever component is added. In Mrk~205, there is no
evidence for relativistic broadening of any emission line, but in Mrk~509
the best fit parameters for an ionised Fe emission line suggest that it 
might originate
in the inner regions of an accretion disk. However, illumination of distant,
cold material provides a simpler, self consistent explanation of the spectral
features than models including reflection from highly ionized, relativistic 
material. Therefore, contrary to Pounds \etal \nocite{pounds01} (2001) and 
Reeves \etal \nocite{reeves01b} (2001b), 
we do not find strong evidence in either
object for
reprocessing in the highly ionised inner parts of an accretion disk. 

\section{Acknowledgments}
Based on observations obtained with \xmm, an ESA science mission with
instruments and contributions directly funded by ESA Member states and 
the USA (NASA).

\begin{appendix}
\section{Method to combine spectra from EPIC MOS and EPIC pn cameras}
\label{sec:appendix}

The following method was designed with pulse height spectra and response 
matrices in
standard `OGIP'\footnote
{http://heasarc.gsfc.nasa.gov/docs/heasarc/ofwg/\\ 
docs/spectra/ogip\_92\_007/ogip\_92\_007.html}
 format in mind.
We use the term `response matrix' to refer
to the product of the effective area and the energy$\to$channel redistribution
 matrix. 

We wish to combine a number of individual source+background pulse-height
spectra to a single
pulse-height spectrum. We label each individual spectrum with the index
$s=1,2,...N_{spec}$, and we use the index $i=1,2,...N_{chan}(s)$ to
denote the channels of each spectrum. Each channel is assigned a `nominal'
energy range of $ENOM_{min}(i)<E<ENOM_{max}(i)$. The number of photons in each
channel $i$ of spectrum $s$ is $C(s,i)$. 
In general, each source+background
spectrum $C(s,i)$ has a corresponding background spectrum $B(s,i)$, with a 
scaling 
factor $F(s)$ relating the geometric area and/or exposure times of the two 
spectra. Each of the original spectra has a
corresponding response matrix, whose elements contain the effective area for a
given channel and a given energy range. The element of the response matrix for
spectrum $s$ corresponding to a particular channel $i$ and a particular energy
range $E_{min}(j)<E<E_{max}(j)$ (where $j=1,2,...N_{range}(s)$) is denoted
$R(s,i,j)$. Throughout we use capitalised indices when refering to the summed
spectrum and response matrix; the index $s$ is omitted when referring to a
combined spectrum or response matrix.

We
define the fractional overlap of channel $i$ of spectrum $s$ with channel $I$
of the combined spectrum to be
\[{\rm if} \ \  ENOM_{min}(s,i) < ENOM_{max}(I)
\]
\[{\rm and} \ \  ENOM_{max}(s,i) > ENOM_{min}(I)
\]
\[\ \ \ f(s,i,I) = \]
\[
\ \ \ \{min[ENOM_{max}(s,i),ENOM_{max}(I)] -
\]
\[
\ \ \ max[ENOM_{min}(s,i),
ENOM_{min}(I)]\}\ /
\]
\[\ \ \ \{ENOM_{max}(s,i)-ENOM_{min}(s,i)\}
\]
\[
{\rm otherwise}
\]
\begin{equation}
\ \ \ f(s,i,I) = 0
\end{equation}
The combined spectrum is constructed by summing all the counts from all the 
spectra in the nominal energy range of each channel. So,
\begin{equation}
C(I) = \sum_{s=1}^{N_{spec}}
\sum_{i=1}^{N_{chan}(s)} f(s,i,I) C(s,i)
\end{equation}
If the nominal energy ranges of the channels of the different spectra do not
coincide with those of the combined spectrum, some randomization of photons
will be required to ensure that the channels of the output spectrum contain
integer numbers of counts.

We combine the individual 
background spectra into a single background spectrum
with a single scaling factor $F$. The following scheme produces a background
photon spectrum with the signal to noise ratio propogated from the individual
background spectra. 
\begin{equation}
B(I) = \frac{1}{F}
\sum_{s=1}^{N_{spec}}
F(s)
\sum_{i=1}^{N_{chan}(s)}
B(s,i)
\end{equation}
where
\begin{equation}
F=\frac
{
\sum_{s=1}^{N_{spec}}
F^{2}(s)
\sum_{i=1}^{N_{chan}(s)}
B(s,i)
}
{
\sum_{s=1}^{N_{spec}}
F(s)
\sum_{i=1}^{N_{chan}(s)}
B(s,i)
}
\end{equation}

The response matrices are easily combined without any complicated weighting, 
provided that the pulse height spectra are all realisations of a single
spectrum (e.g. if they are from observations of the same source at the same
time but by different instruments). If the individual pulse height spectra are
realisations of {\em intrinsically different} spectra, then the response 
matrix combination
described here will not generally be appropriate. However if the spectra differ
only in intensity then a simple scaling factor 
can be used to weight the
contributions of the different response matrices to the final spectrum. 
For each energy range of the combined response matrix, the elements 
corresponding to
each channel are combined in the same way as the original spectra. This is more
complicated if the energy ranges differ between the response matrices. In this
case we use a weighted average response for a given energy range.
We define the fractional overlap of energy range $j$ of spectrum $s$ with the
energy range $J$ of the output response matrix as follows:
\[{\rm if} \ \  E_{min}(s,j) < E_{max}(J)
\]
\[{\rm and} \ \  E_{max}(s,j) > E_{min}(J)
\]
\[\ \ \ g(s,j,J) =\]
\[\ \ \ \frac{min[E_{max}(s,j),E_{max}(J)]-max[E_{min}(s,j),
E_{min}(J)]}{E_{max}(s,j)-E_{min}(s,j)}
\]
\[
{\rm otherwise}
\]
\begin{equation}
\ \ \ g(s,j,J) = 0 
\end{equation}
The combined response matrix is then constructed according to:
\[R(I,J) = \]
\begin{equation}
\sum_{s=1}^{N_{spec}}
\sum_{i=1}^{N_{chan}(s)}
f(s,i,I)
\frac{
\sum_{j=1}^{N_{range}(s)} g(s,j,J) R(s,i,j)
}{
\sum_{j=1}^{N_{range}(s)} g(s,j,J)
}
\end{equation}

\end{appendix}

\end{document}